\documentclass[11pt,aps,superscriptaddress,preprintnumbers,amsmath,amssymb,nofootinbib]{revtex4}
\usepackage{epsfig}  
\usepackage{graphicx}
\usepackage{hyperref}

\hypersetup{
    colorlinks=true,
    citecolor=red,
    linkcolor=blue,
    filecolor=green,      
    urlcolor=magenta,
}
\usepackage{color}
\usepackage{float}
\usepackage{amsfonts}
\usepackage{amsmath}
\usepackage{slashed}
\usepackage{soul}
\large

\begin{document} 

\title{Improved Bounds and Global Fit of Flavor-Violating Charged Lepton Yukawa Couplings post LHC}

\author{Fayez Abu-Ajamieh\footnote{Currently at  SOKN Engineering, Whitefish, Montana, USA.}}\email{fayezabuajamieh@gmail.com}
\affiliation{Centre for High Energy Physics, Indian Institute of Science, Bangalore 560012, India}

\author{Suman Kumbhakar}
\email{kumbhakar.suman@gmail.com}
\affiliation{Department of Physics, University of Calcutta, 92 Acharya Prafulla Chandra Road, Kolkata 700009, India}

\author{Ratan Sarkar}
\email{ratansarkar@imsc.res.in}
\affiliation{The Institute of Mathematical Sciences, CIT Campus, Taramani Chennai 600113, Tamil Nadu, India}

\author{Sudhir Vempati}
\email{vempati@iisc.ac.in}
\affiliation{Centre for High Energy Physics, Indian Institute of Science, Bangalore 560012, India}

\begin{abstract}
 Higgs couplings to charged leptons form an important measurement to understand not only the Standard Model (SM), but also physics Beyond Standard Models (BSM). In this work, we update the bounds on the Flavor-Violating (FV) Higgs couplings to charged leptons. We find that the bounds on the size of the couplings could range between $\sim \mathcal{O}(10^{-3}) - \mathcal{O}(10^{-6})$. In fact, the direct constraints from LHC are much stronger than those inferred indirectly from rare decays in the $\tau-\mu$ and $\tau -e $ sector.  We also match these bounds to the SM Effective Field Theory (SMEFT) and find lower limits on the scale of New Physics (NP). We find that the scale of NP ranges between $\sim \mathcal{O}(10) - \mathcal{O}(100)$ TeV. We also present future projections for some upcoming experiments. We find that the current bounds on the couplings to $\mu- e$ are stronger than all future projections. 
 
 \end{abstract}
\maketitle

\section{Introduction}
The discovery of the Higgs boson at 125 GeV in the LHC \cite{ATLAS:2012yve, CMS:2012qbp} represented a key milestone in finalizing the SM of particle physics. Ever since, the LHC has reported on the properties of the discovered boson \cite{ATLAS:2016neq, ATLAS:2017azn, CMS:2022dwd, ATLAS:2022vkf}, revealing excellent agreement with the predictions of the SM. Nonetheless, the level of experimental measurements still allows for significant deviation from the SM predictions, allowing for BSM physics \cite{Abu-Ajamieh:2020yqi, Abu-Ajamieh:2021vnh, Abu-Ajamieh:2021egq, Abu-Ajamieh:2022ppp, Abu-Ajamieh:2022dtm, Dawson:2022zbb, Abu-Ajamieh:2022nmt}. Some of the BSM proposals include FV in the Higgs sector. It is known that in the SM, the Higgs boson does not violate flavor, however, it is quite easy to generalize the Higgs couplings to include FV by assuming NP at high energy. Studying FV in the Higgs sector has a long history \cite{Bjorken:1977vt, McWilliams:1980kj, Shanker:1981mj, Barr:1990vd, Babu:1999me, Diaz-Cruz:1999sns, Han:2000jz, Blanke:2008zb, Casagrande:2008hr, Giudice:2008uua, Aguilar-Saavedra:2009ygx, Albrecht:2009xr, Buras:2009ka, Agashe:2009di, Goudelis:2011un, Arhrib:2012mg, McKeen:2012av, Azatov:2009na, Blankenburg:2012ex, Kanemura:2005hr, Davidson:2010xv, DAmbrosio:2002vsn} (for FV in the $Z$ boson decays see \cite{Brignole:2004ah, Davidson:2012wn, Goto:2015iha, Kamenik:2023hvi, Jueid:2023fgo, Abu-Ajamieh:2025vxw}). For instance, it was argued in ref.~\cite{DAmbrosio:2002vsn} that the lightest quarks in the SM quark sector have an approximate $U(1)^{5}$ symmetry that is broken by the Yukawa couplings, which suggests that flavor does not have to be conserved in the SM. Given that any FV is excluded at the $\sim \mathcal{O}(\text{TeV})$ scale led the authors to suggest Minimal Flavor Violation (MFV), where all FV arises from the Yukawa interactions only. MFV was thoroughly investigated in ref.~\cite{Harnik:2012pb}, where bounds on the FV Yukawa couplings of the quarks and leptons were extracted from various experimental measurements. The same argument was extended to di-Higgs FV interactions in ref.~\cite{Abu-Ajamieh:2024rzw}. 
In this paper, we seek to update the experimental bounds of ref.~\cite{Harnik:2012pb} and set further bounds on the FV Higgs Yukawa couplings. Some of the experimental measurements have improved, making the bounds more stringent. Lepton FV Higgs couplings can arise in several well-motivated BSM frameworks, such as the two-Higgs doublet model (2HDM)~\cite{Branco:2011iw}, R-partiy violating minimal supersymmetric SM (MSSM)~\cite{Arhrib:2012ax}, leptoquark-mediated interactions in Grand Unified Theories (GUTs)~\cite{Dorsner:2016wpm}, certain $Z'$ models~\cite{Langacker:2000ju}, models with vector-like leptons~\cite{Falkowski:2013jya}, and models with warped extra dimensions \cite{Agashe:2004cp}. These frameworks can generate such couplings at tree or loop level, making the derived bounds valuable probes of NP in a broad class of BSM models.

Following a bottom-up approach, it is quite easy to introduce MFV in the Higgs sector after Electroweak Symmetry Breaking (EWSB) by writing the effective Lagrangian of the Yukawa interaction as follows
\begin{equation}\label{eq:FV_Lag}
    \mathcal{L_{\text{Y}}} = -m_{i}\bar{f}_{L}^{i}f_{R}^{i} - Y_{ij}\bar{f}_{L}^{i}f_{R}^{j}  h  + h.c. + \cdots,
\end{equation}
where $Y_{ij}$ are now promoted to be non-diagonal complex matrices. It is easy to see that the SM would correspond to the case where the matrices are diagonal $Y_{ij} = (m_{i}/v)\delta_{ij}$. Such deviations could stem from higher-dimensional operators arising from integrating out heavy degrees of freedom. Consider for instance the six-dimensional operator
\begin{equation}\label{eq:dim6_operator}
\mathcal{L}_{6} = \frac{\lambda_{L}^{ij}}{\Lambda^{2}}(\bar{f}_{L}^{i}\gamma^{\mu}f_{L}^{j})(H^{\dagger}i\overleftrightarrow{D}_{\mu}H) + \frac{\lambda_{R}^{ij}}{\Lambda^{2}}(\bar{f}_{R}^{i}\gamma^{\mu}f_{R}^{j})(H^{\dagger}i\overleftrightarrow{D}_{\mu}H) -\Bigg[\frac{\lambda_{ij}}{\Lambda^{2}}(\bar{f}_{L}^{i}f_{R}^{j})H(H^{\dagger}H) + h.c. \Bigg],
\end{equation}
where $\Lambda$ is the scale of NP, $(H^{\dagger}i\overleftrightarrow{D}_{\mu}H) \equiv H^{\dagger}iD_{\mu}H - (iD_{\mu}H^{\dagger})H$, and the Wilson coefficients $\lambda_{ij}$, $\lambda_{L,R}^{ij}$ could be non-diagonal complex matrices. After EWSB and after diagonalizing the mass matrices, these operators will lead to non-diagonal contributions to the Yukawa matrices
\begin{equation}\label{eq:Yukawa_tota}
    Y_{ij} = \frac{m_{i}}{v}\delta_{ij} + \frac{v^{2}}{\sqrt{2}\Lambda^{2}} \hat{\lambda}_{ij},
\end{equation}
where $\hat{\lambda}_{ij} = U_{L}\lambda_{ij}U^{\dagger}_{R}$, with $U$ being the unitary matrix that diagonalizes the mass matrix. Notice here that the higher-order operators will also contribute to the diagonal elements that correspond to the SM Yukawa couplings. Thus, to remain within experimental limits, such diagonal contributions need to be suppressed compared to the non-diagonal ones. This involves a certain level of fine-tuning of any UV completion which is unavoidable barring any hidden symmetry that forces such a suppression. Also note that $Y_{ij}$ need not be symmetric, and in fact, they were treated as non-symmetric in ref.~\cite{Harnik:2012pb}.

In this paper, we limit ourselves to the lepton sector. We delay treating the quark sector to future work. We review the experimental bounds on the FV Yukawa couplings arising from the LHC direct searches, FV lepton decays $\tau \rightarrow \mu \gamma$, $\tau \rightarrow e \gamma$ and $\mu \rightarrow e \gamma$; lepton number-violating decays $\tau \rightarrow 3\mu$, $\tau \rightarrow 3e$ and $\mu \rightarrow 3 e$; the anomalous magnetic dipole moments of the muon and electron $(g-2)_{\mu,e}$; the electric dipole moments of the muon and electron $\text{EDM}_{\mu,e}$; muonium-antimuonium oscillation and from muon conversion in nuclei. Some of the bounds, in particular those from LHC  represent improvement over those found in ref.~\cite{Harnik:2012pb}. We also convert these bounds on the Yukawa couplings to lower limits on the scale of NP by utilizing the SMEFT. We find that the scale of NP ranges between $\sim \mathcal{O}(10) - \mathcal{O}(100)$ TeV, depending on the coupling. We perform a $\chi^2$ analysis to determine the size of each FV Yukawa coupling. In order to do that, we assume these couplings to be complex. We also present the allowed regions for different confidence intervals.

This paper is organized as follows. In section II, we discuss the bounds from direct and indirect searches. In section III, we describe the $\chi^2$ fit and fit results. In section IV, we evaluate the scale of NP using matching to the SMEFT. In section V, we present our conclusions.

\section{Bounds on the FV Higgs Yukawa Couplings to Leptons}\label{Sec:2}
In this Section, we update the experimental bounds on the FV Yukawa couplings of charged leptons, derived from the latest relevant experimental searches. In our approach, we mostly follow a similar argument as in ref.~\cite{Harnik:2012pb}, however, in addition to the updated bounds, there are several novelties in our work, including matching to the SMEFT. Explicitly, the Lagrangian describing FV Higgs decays in the lepton sector can be written as
\begin{equation}\label{eq:Explicit_lag}
    \mathcal{L}_{\text{LFV}} \supset -Y_{e\mu}\bar{e}_L\mu_R h -Y_{\mu e}\bar{\mu}_L e_R h -Y_{e\tau}\bar{e}_L
\tau_R h -Y_{\tau e}\bar{\tau}_L e_R h - Y_{\mu\tau}\bar{\mu}_L \tau_R h -Y_{\tau\mu}\bar{\tau}_L \mu_R h + h.c.,
\end{equation}
where in the most general case $Y_{ij} \neq Y_{ji}$. We set the diagonal Yukawa couplings equal to the SM values and extract bounds on the FV off-diagonal elements. All bounds are discussed below in details and summarized in Table \ref{tab:1}.

\begin{table}
\centering
 \resizebox{\textwidth}{!}{ 
\begin{tabular}{c|c|c|c|c|c}
\hline \hline
Sources & Couplings & Bounds from ref.~\cite{Harnik:2012pb}& Bounds from this work & $\Lambda$ (TeV) & Future Projections \\ 
\hline 
$\tau \rightarrow \mu \gamma $  &$\sqrt{|Y_{\tau \mu}|^2+|Y_{\mu \tau}|^2}$ &  $ 1.6\times 10^{-2}$  &$ 1.6\times 10^{-2}$  & $2.4$ & $2.4\times 10^{-3}$\\ 
\hline 
$\tau \rightarrow 3 \mu $ & $\sqrt{|Y_{\tau \mu}|^2+|Y_{\mu \tau}|^2}$ & 0.25  &$ 0.25$  & $0.6$ & $0.038$\\ 
\hline 
$(g-2)_{\mu}$ & {\rm Re}$(Y_{\mu \tau}Y_{\tau \mu })$ &  $(2.7\pm 0.75)\times 10^{-3}$ &$(2.33\pm 0.45)\times 10^{-3}$ & $1.1 - 1.2$ & $-$\\ 
\hline 
 $\rm EDM_{\mu}$ &  ${\rm Im}(Y_{\mu \tau}Y_{\tau \mu })$&  $[-0.8,1.0]$ & $[-0.8, 1.0]$ & $0.3$ &$-$ \\ 
 \hline
\hline
$h\to \mu\tau$ & $\sqrt{|Y_{\tau \mu}|^2+|Y_{\mu \tau}|^2}$ & $-$ & $1.06\times 10^{-3}$ & $9.3$ & $4.72\times 10^{-4}$\\ 
\hline
\hline 
$\tau \rightarrow e \gamma $ & $\sqrt{|Y_{\tau e}|^2+|Y_{e \tau}|^2}$ & $1.40\times 10^{-2}$  & $1.38\times 10^{-2}$ & $2.6$ & $2.4\times 10^{-3}$ \\ 
\hline
$\tau \rightarrow 3 e$ & $\sqrt{|Y_{\tau e}|^2+|Y_{e \tau}|^2}$ & $0.12$ & $ 0.12$  & $0.9$ & $0.017$\\ 
\hline 
$(g-2)_e$ & ${\rm Re}(Y_{e \tau}Y_{\tau e })$ &  $[-2.1, 2.9]\times 10^{-3}$& $(6.61\pm 3.18)\times 10^{-5}$ & $2.5 - 3.3$ & $-$\\ 
\hline 
$\rm EDM_e$ &  ${\rm Im}(Y_{e \tau}Y_{\tau e })$& $1.1\times 10^{-8}$ & $[-0.80,3.42]\times 10^{-11}$ & $ 105$ & $-$\\
\hline
\hline
$h\to e\tau$ &  $\sqrt{|Y_{\tau e}|^2+|Y_{e \tau}|^2}$ & $-$ & $1.22\times 10^{-3}$ & $8.6$ & $4.72\times 10^{-4}$\\ 
\hline
\hline 
$\mu \rightarrow e \gamma$   &$\sqrt{|Y_{ \mu e}|^2+|Y_{e \mu }|^2}$ & $3.6\times 10^{-6}$ &$2.5 \times 10^{-5} $ & $60$ & $6.22\times 10^{-6}$\\ 
\hline 
$\mu \rightarrow 3e $ & $\sqrt{|Y_{ \mu e}|^2+|Y_{e \mu }|^2}$ & $3.1\times 10^{-5}$ &$3.1 \times 10^{-5}$  & $54$ & $3.1\times 10^{-7}$ \\ 
\hline 
$\rm EDM_e$ &  ${\rm Im}(Y_{e \mu}Y_{\mu e })$ & $9.8\times 10^{-8}$ & $[-0.75, 3.18]\times 10^{-10}$ & $60$ & $-$\\
\hline 
$(g-2)_e$  &${\rm Re}(Y_{e \mu}Y_{\mu e })$ & $[-0.019, 0.026]$& $(6.16\pm 2.96)\times 10^{-4}$ & $1.5 - 1.9$ & $-$\\ 
\hline 
$M-\overline{M}$ oscillation & $|Y_{\mu e}+Y_{e\mu}^*|$ & $0.079$ & $0.079$ & $0.7$  & $-$\\ 
\hline 
$\mu \rightarrow e$ conversion & $\sqrt{|Y_{ \mu e}|^2+|Y_{e \mu}|^2}\,$ & $1.2\times 10^{-5}$ & $1.2 \times 10^{-5}$  & $87$ & $1.71\times 10^{-9}$\\ 
\hline
\hline
$h\to \mu e$ & $\sqrt{|Y_{\mu e}|^2+|Y_{e \mu}|^2}$ & $-$ & $1.81\times 10^{-4}$ & $22$ & $1.49\times 10^{-4}$\\ 
\hline \hline
$\mu \rightarrow e \gamma$ & $(|Y_{\tau \mu}Y_{e\mu}|^2+|Y_{\mu \tau}Y_{\tau e}|)^\frac{1}{4}$ & $3.4\times 10^{-4}$ & $ 2.18 \times 10^{-4}$  & $20$ & $1.22\times 10^{-4}$\\
\hline\hline
\end{tabular}
}
\caption{Current limits on the FV Yukawa couplings from the relevant observables. For comparison, we also list the earlier bounds reported in ref.~\cite{Harnik:2012pb}. The corresponding lower limits on the NP scale inferred from the current bounds are indicated for each case. Future projected bounds are also listed.}
\label{tab:1}
\end{table}

\subsection{Bounds from Direct Searches}
The simplest and most straightforward bounds are obtained from direct searches of FV leptonic Higgs decays from the LHC. These decays proceed at tree-level with the decay width given by
\begin{equation}\label{eq:FV_Decay}
    \Gamma(h \rightarrow \bar{\ell}_{i}\ell_{j}) \sim \frac{m_{h}}{8\pi}(|Y_{ij}|^{2} + |Y_{ji}|^{2}),
\end{equation}
where $\ell_{i,j} = \{\tau, \mu ,e\}$. The current bound on $h \rightarrow \tau\mu$ from CMS is given by $1.5 \times 10^{-3}$ at $95\%$ CL~\cite{CMS:2021rsq}, which translates to $\sqrt{|Y_{\tau\mu}|^{2}+|Y_{\mu\tau}|^{2}} < 1.06 \times 10^{-3}$, whereas bound on $h \rightarrow \tau e$ is given by $2 \times 10^{-3}$ at $95\%$ CL \cite{ParticleDataGroup:2024cfk}, which translates to $\sqrt{|Y_{\tau e}|^{2}+|Y_{ e\tau}|^{2}} < 1.22 \times 10^{-3}$. Finally, the bound on $h \rightarrow \mu e$ also from CMS is given by $4.4 \times 10^{-5}$ at $95\%$ CL \cite{CMS:2023pte}, which gives $\sqrt{|Y_{\mu e}|^{2}+|Y_{ e\mu}|^{2}} < 1.81 \times 10^{-4}$. Better bounds are expected from the high luminosity run of the LHC (HL-LHC)~\cite{Cepeda:2019klc}. Specifically, the projected bounds on $\text{Br}(h \rightarrow \tau \mu (e)) < 3\times 10^{-4}$ translate to $\sqrt{|Y_{\tau\mu(e)}|^{2}+|Y_{\mu\tau(e)}|^{2}} < 4.72 \times 10^{-4}$, whereas the projected bound on $\text{Br}(h \rightarrow \mu e) < 3\times 10^{-5}$ translates to $\sqrt{|Y_{\mu e}|^{2}+|Y_{\mu e}|^{2}} < 1.49 \times 10^{-4}$. In our calculation, we have set decay width of Higgs $\Gamma_{h} = 3.7$ MeV.

\subsection{Constraints from $\tau \rightarrow \mu\gamma$, $\tau \rightarrow e\gamma$ and $\mu \rightarrow e\gamma$}
The decay $\tau \rightarrow \mu \gamma$ proceeds at loop-level. It was shown in ref.~\cite{Harnik:2012pb} that the 2-loop contribution is as important as the 1-loop contribution. The 1-loop and 2-loop contributions can be found in Figures 1 and 12 in ref.~\cite{Harnik:2012pb}, and the decay width can be expressed as 
\begin{equation}\label{eq:tau_to_mu_gamma}
    \Gamma(\tau \rightarrow \mu \gamma) = \frac{\alpha m_{\tau}^{5}}{64\pi^{5}}(|c_{L}|^{2} + |c_{R}|^{2}), 
\end{equation}
where $c_{L}$ and $c_{R}$ are the Wilson coefficients, which at one loop are given by
\begin{equation}\label{eq:Wilson_1loop1}
    c_{L}^{\text{1loop}} \simeq \frac{1}{12m_{h}^{2}}Y_{\tau\tau}Y^{*}_{\tau\mu}\Big(-4 +3\log\frac{m_{h}^{2}}{m_{\tau}^{2}} \Big), \hspace{5mm} c_{R}^{\text{1loop}} \simeq \frac{1}{12m_{h}^{2}}Y_{\mu\tau}Y_{\tau\tau}\Big(-4 +3\log\frac{m_{h}^{2}}{m_{\tau}^{2}} \Big),
\end{equation}
and at two loops
\begin{equation}\label{eq:Wilson_2loop}
    c_{L}^{\text{2loop}} \simeq \frac{0.055Y^{*}_{\tau\mu}}{(125\text{GeV})^{2}}, \hspace{5mm} c_{R}^{\text{2loop}} \simeq \frac{0.055Y_{\mu\tau}}{(125\text{GeV})^{2}}.
\end{equation}

The decay widths for $\mu \rightarrow e\gamma$ and $\tau \rightarrow e\gamma$ are obtained with replacing $\tau \rightarrow \mu$ and $\mu \rightarrow e$ in the equations for the first, and by $\mu \rightarrow e$ for the second. The latest bound on the first decay reads $\text{Br}(\tau \rightarrow \mu \gamma)< 4.2 \times 10^{-8}$ \cite{Belle:2021ysv}, which translates to $\sqrt{|Y_{\tau\mu}|^{2} + |Y_{\mu\tau}|^{2}} < 1.6 \times 10^{-2}$. The bound on second decay reads $\text{Br}(\tau \rightarrow e \gamma)< 3.3 \times 10^{-8}$ \cite{BaBar:2009hkt}, which translates to $\sqrt{|Y_{\tau e}|^{2} + |Y_{e \tau}|^{2}} < 1.38 \times 10^{-2}$. Finally, the bound on the last decay is given by $\text{Br}(\mu \rightarrow e \gamma)< 4.2 \times 10^{-13}$ \cite{MEG:2016leq}, which translates to $\sqrt{|Y_{\mu e}|^{2} + |Y_{e \mu}|^{2}} < 2.5 \times 10^{-5}$. It should also be noted that the decay $\mu \rightarrow e\gamma$ can be used to set bound on the combination $Y_{\mu\tau}Y_{\tau e}$ by using the 1-loop Wilson coefficients
\begin{equation}\label{eq:Wilson_1loop2}
    c_{L}^{\text{1loop}} \simeq \frac{1}{8m_{h}^{2}} \frac{m_{\tau}}{m_{\mu}}Y^{*}_{\mu\tau}Y^{*}_{\tau e}\Big(-4 +3\log\frac{m_{h}^{2}}{m_{\tau}^{2}} \Big), \hspace{5mm} c_{R}^{\text{1loop}} \simeq  \frac{1}{8m_{h}^{2}} \frac{m_{\tau}}{m_{\mu}}Y_{\tau\mu}Y_{e \tau}\Big(-4 +3\log\frac{m_{h}^{2}}{m_{\tau}^{2}} \Big).
\end{equation}

The experimental measurement yields the bound $(|Y_{\tau\mu}Y_{e\tau}|^{2} + Y_{\mu\tau}Y_{\tau e}|^{2})^{1/4} < 2.18\times 10^{-4}$. All these bounds are given at $90\%$ CL. Future projections could give more stringent bounds on the Yukawa couplings. With $50$ ab$^{-1}$ of data to be accumulated at SuperKEKB, the future sensitivity could reach Br($\tau\to \{\mu,e\}\gamma)<1.0\times 10^{-9}$~\cite{Aushev:2010bq}. This sets the projected bounds to be $\sqrt{|Y_{\tau \mu (e)}|^{2} + |Y_{ \mu \tau (e)}|^{2}} < 2.4 \times 10^{-3}$. The future sensitivity of Br($\mu\to e\gamma$) is $4.0\times 10^{-14}$ from the MEG-II experiment~\cite{Renga:2014xra}. This leads to the projected bounds $\sqrt{|Y_{e \mu}|^{2} + |Y_{ \mu e}|^{2}} < 6.22 \times 10^{-6}$ and $(|Y_{\tau\mu}Y_{e\tau}|^{2} + Y_{\mu\tau}Y_{\tau e}|^{2})^{1/4} < 1.22\times 10^{-4}$.

\subsection{Constraints from $\tau \rightarrow 3\mu$, $\tau \rightarrow 3 e$, $\mu \rightarrow 3 e$}
Lepton number violating decays could proceed either at tree level through a Higgs mediator, or at loop level via the loop diagrams from the above decay after integrating out the loops, with the photon decaying to two identical leptons. The decay width of $\tau \rightarrow 3\mu$ is given by
\begin{equation}\label{eq:tau_to_3mu}
    \Gamma(\tau \rightarrow 3\mu) \simeq \frac{\alpha m_{\tau}^{5}}{6(2\pi)^{5}}\Big| \log{\frac{m_{\mu}^{2}}{m_{\tau}^{2}}} - \frac{11}{4}\Big|(|c_{L}|^{2} + |c_{R}|^{2}),
\end{equation}
where the Wilson coefficients are given by eqs.(\ref{eq:Wilson_1loop1}) and (\ref{eq:Wilson_2loop}) above. The decays $\tau \rightarrow 3e$ and $\mu \rightarrow 3e$ can be obtained with the same replacements described above. The latest experimental measurements are given by $\text{Br}(\tau \rightarrow 3\mu) < 2.1 \times 10^{-8}$ \cite{Hayasaka:2010np}, $\text{Br}(\tau \rightarrow 3e) < 2.7 \times 10^{-8}$ \cite{Hayasaka:2010np} and $\text{Br}(\mu \rightarrow 3e) < 1 \times 10^{-12}$ \cite{ParticleDataGroup:2024cfk} at $90\%$ CL, which translate to $\sqrt{|Y_{\tau\mu}|^{2} + |Y_{\mu\tau}|^{2}} < 0.25$, $\sqrt{|Y_{\tau e}|^{2} + |Y_{e \tau}|^{2}} < 0.12$ and $\sqrt{|Y_{\mu e}|^{2} + |Y_{e \mu}|^{2}} < 3.1 \times 10^{-5}$ respectively. The future projections on Br$(\tau\to 3\mu)$ and Br$(\tau\to 3e)$ are given by $5\times 10^{-10}$ from the Belle-II experiment~\cite{Aushev:2010bq,Belle-II:2022cgf}. This leads to the projected bounds $\sqrt{|Y_{\tau\mu}|^{2} + |Y_{\mu\tau}|^{2}} < 0.038$ and $\sqrt{|Y_{\tau e}|^{2} + |Y_{e\tau}|^{2}} < 0.017$, respectively. On the other hand, the future limit on Br$(\mu \to 3e) < 10^{-16}$ taken also from the Belle II experiment~\cite{Banerjee:2022vdd,Calibbi:2017uvl} leads to the projected bound $\sqrt{|Y_{\mu e}|^{2} + |Y_{e \mu}|^{2}} < 3.1 \times 10^{-7}$.

\subsection{Constraints from Muonium-antimuonium Oscillation}
Muonium (a bound state of $\mu^{+}e^{-}$) can oscillate to an antimuonium (a bound state of $\mu^{-}e^{+}$), with the time-integrated conversion probability given by \cite{Harnik:2012pb}
\begin{equation}\label{eq:ConversionP}
    P(M \rightarrow \overline{M}) = \int_{0}^{\infty}dt \Gamma_{\mu}\sin^{2}(\Delta M t)e^{-\Gamma_{\mu}t} = \frac{2}{\Gamma^{2}_{\mu}/(\Delta M)^{2}+4},
\end{equation}
where $\Gamma_{\mu}$ is the muon decay width, and $\Delta M$ is the mass splitting between the two states \cite{Clark:2003tv}
\begin{equation}
    \Delta M = \frac{|Y_{\mu e} + Y_{e \mu}^{*}|^{2}}{2\pi a^{3} m_{h}^{2}},
\end{equation}
where $ a = (m_{e}+m_{\mu})/(m_{e}m_{\mu}\alpha)$ is the muonium Bohr radius. The time-integrated conversion probability is constrained by the MACS experiment at PSI \cite{Willmann:1998gd} to be $P(M \rightarrow \overline{M}) < 8.3 \times 10^{-11}/S_{B}$, where $S_{B}$ accounts for the splitting of the muonium states in the magnetic field of the detector. Taking $S_{B} = 0.35$ as in ref.~\cite{Harnik:2012pb}, we obtain the bound $|Y_{\mu e} + Y_{e \mu}^{*}| < 0.079$ at $90\%$ CL, which is the same as in ref.~\cite{Harnik:2012pb}.
\subsection{The Magnetic Dipole Moments}
FV Higgs couplings to $\tau\mu$ contribute to the magnetic dipole moment of the muon at 1-loop. This contribution is given by \cite{Harnik:2012pb,Blankenburg:2012ex}\footnote{Notice that the Higgs coupling to $\mu e$ also contributes to $(g-2)_{\mu}$, however, this contribution is suppressed by the small mass of the electron.}

\begin{equation}\label{eq:g-2}
    a_{\mu} \equiv \frac{(g-2)_{\mu}}{2} \simeq \frac{\text{Re}(Y_{\mu\tau}Y_{\tau\mu})}{8\pi^{2}}\frac{m_{\mu}m_{\tau}}{2m_{h}^{2}}\Big( 2\log{\frac{m_{h}^{2}}{m_{\tau}^{2}}}-3\Big).
\end{equation}
The discrepancy between the SM predictions and experimental measurements stands at \cite{Muong-2:2021ojo, Muong-2:2023cdq}\footnote{We should point out that the recent high-precision lattice gauge QCD simulations \cite{Borsanyi:2020mff, Ce:2022kxy, ExtendedTwistedMass:2022jpw} appear to be closer to the experimental results, thereby lowering the significance of the gap, however, for the purposes of this paper, we assume that the discrepancy is real and is given by eq. (\ref{eq:mu_discpreancy}).}
\begin{equation}\label{eq:mu_discpreancy}
    \Delta a_{\mu} \equiv a_{\mu}^{\text{Exp}} - a_{\mu}^{\text{SM}} = (249 \pm 48) \times 10^{-11},
\end{equation}
which requires $\text{Re}(Y_{\mu\tau}Y_{\tau\mu}) = (2.33 \pm 0.45) \times 10^{-3}$ to solve it. It should also be noted that FV Higgs couplings could also contribute to the magnetic dipole moment of the electron. The magnetic dipole moment for the electron can be obtained by replacing $\mu \rightarrow e$ in eq.  (\ref{eq:g-2}) to obtain bound on the combination $Y_{\tau e} Y_{e\tau}$, and by replacing $\tau \rightarrow \mu$ in the same equation to obtain bound on the combination $Y_{\mu e} Y_{e\mu}$. The latest measurement of the electron magnetic dipole moment reveals the following discrepancy with the SM prediction \cite{Fan:2022eto}
\begin{equation}\label{eq:e_discpreancy}
    \Delta a_{e} \equiv a_{e}^{\text{Exp}} - a_{e}^{\text{SM}} = (3.41 \pm 1.64) \times 10^{-13},
\end{equation}
which requires either $\text{Re}(Y_{e\tau}Y_{\tau e}) = (6.61 \pm 3.18) \times 10^{-5}$ or $\text{Re}(Y_{\mu e}Y_{e\mu}) = (6.16 \pm 2.96) \times 10^{-4}$ to solve it. However, as can be seen from Fig.~\ref{bounds}, these regions are excluded by the other bounds. Thus, the FV Higgs interactions cannot account for either $(g-2)_{\mu}$ or  $(g-2)_{e}$, at least not on their own. 

\subsection{Constraints from Electric Dipole Moments}
If the FV Higgs couplings to leptons are allowed to be complex, then they would contribute to the Electric Dipole Moments (EDM) of the muon and electron. The contribution of the FV Yukawa couplings to the EDM of the muon is given by \cite{Harnik:2012pb}
\begin{equation}\label{eq:EDM}
    d_{\mu} \simeq - \frac{\text{Im}(Y_{\tau\mu}Y_{\mu\tau})}{16\pi^{2}}\frac{e m_{\tau}}{2m_{h}^{2}}\Big(2\log{\frac{m_{h}^{2}}{m_{\tau}^{2}}} -3 \Big),
\end{equation}
and the EDM for the electron can be obtained by replacing $Y_{\tau\mu}Y_{\mu\tau} \rightarrow Y_{\tau e}Y_{e \tau}$ with $\tau$ running in the loop, and by replacing $Y_{\tau\mu}Y_{\mu\tau} \rightarrow Y_{ e\mu}Y_{\mu e}$ and $m_{\tau} \rightarrow m_{\mu}$ with $\mu$ running in the loop. The experimental bound on the muon EDM indicates that $d_{\mu} = (-0.1 \pm 0.9) \times 10^{-19}$ at $\hspace{1mm}95\%$ CL \cite{Muong-2:2008ebm}, which translates into the bound $-0.8 < \text{Im}(Y_{\mu\tau}Y_{\tau\mu}) < 1.0$. On the other hand, the experimental limit on the EDM for the electron is given by $d_{e} = (-1.3 \pm 2.1)\times 10^{-30}$ at $90\%$ CL \cite{Roussy:2022cmp}, which translates to $-0.8\times 10^{-11} < \text{Im}(Y_{\mu\tau}Y_{\tau\mu}) < 3.42 \times 10^{-11}$ for $\tau$ running in the loop, and $-0.75\times 10^{-10} < \text{Im}(Y_{\mu\tau}Y_{\tau\mu}) < 3.18 \times 10^{-10}$ for $\mu$ running in the loop.

\subsection{Constraints from $\mu \rightarrow e$ Conversion in Nuclei}
It is possible to set stringent bounds on the FV Higgs couplings to $\mu$ and $e$ from experiments searching for $\mu \rightarrow e$ conversion in nuclei. The full treatment is provided in ref.~\cite{Harnik:2012pb}. Here, we provide the final results. The decay width of the muon conversion can be expressed as
\begin{align}\label{eq:mu_Conv}
    \Gamma(\mu \rightarrow e) & = \Big| -\frac{e}{16\pi^{2}}c_{R}D + \Tilde{g}_{LS}^{(p)}S^{(p)} + \Tilde{g}_{LS}^{(n)}S^{(n)} + \Tilde{g}_{LV}^{(p)}V^{(p)}\Big|^{2} \nonumber\\
    & + \Big| -\frac{e}{16\pi^{2}}c_{L}D + \Tilde{g}_{RS}^{(p)}S^{(p)} + \Tilde{g}_{RS}^{(n)}S^{(n)} + \Tilde{g}_{RV}^{(p)}V^{(p)}\Big|^{2},
\end{align}
where $c_{L,R}$ are the same Wilson coefficients given in eqs (\ref{eq:Wilson_1loop1}) and (\ref{eq:Wilson_2loop}) and full expressions of the couplings $\Tilde{g}_{L,R;S}^{(n,p)}$, and $\Tilde{g}_{L,R; V}^{(n,p)}$ can be found in ref.~\cite{Harnik:2012pb}. The coefficients $D$, $S^{(p)}$, $S^{(n)}$ and $V^{(p)}$ are the overlap integrals of the muon, electron and nucleon wave functions and are tabulated for various materials in ref.~\cite{Kitano:2002mt}. 

\begin{figure}
    \centering
    \begin{tabular}{cc}
    \includegraphics[width=75mm]{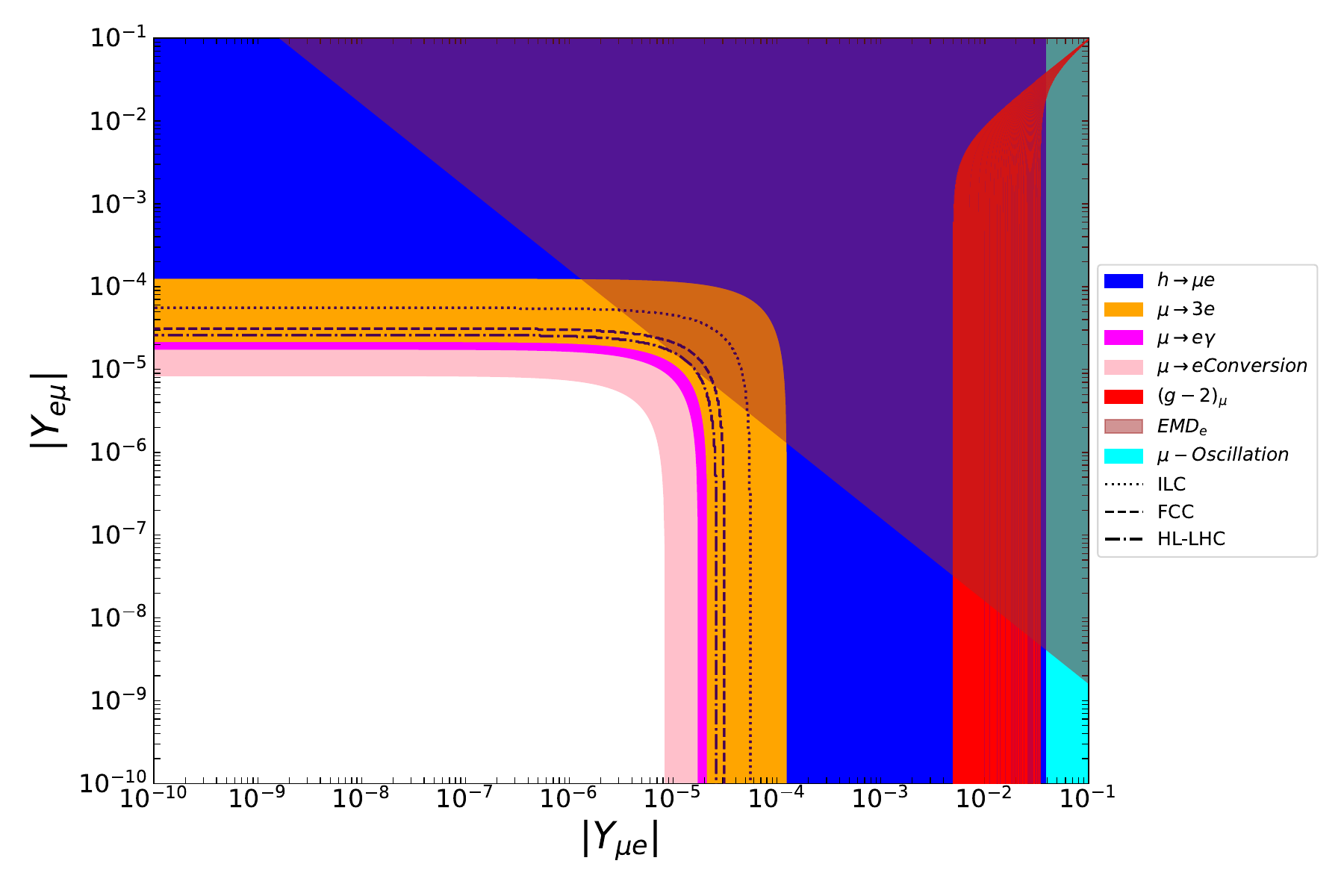} & \includegraphics[width=75mm]{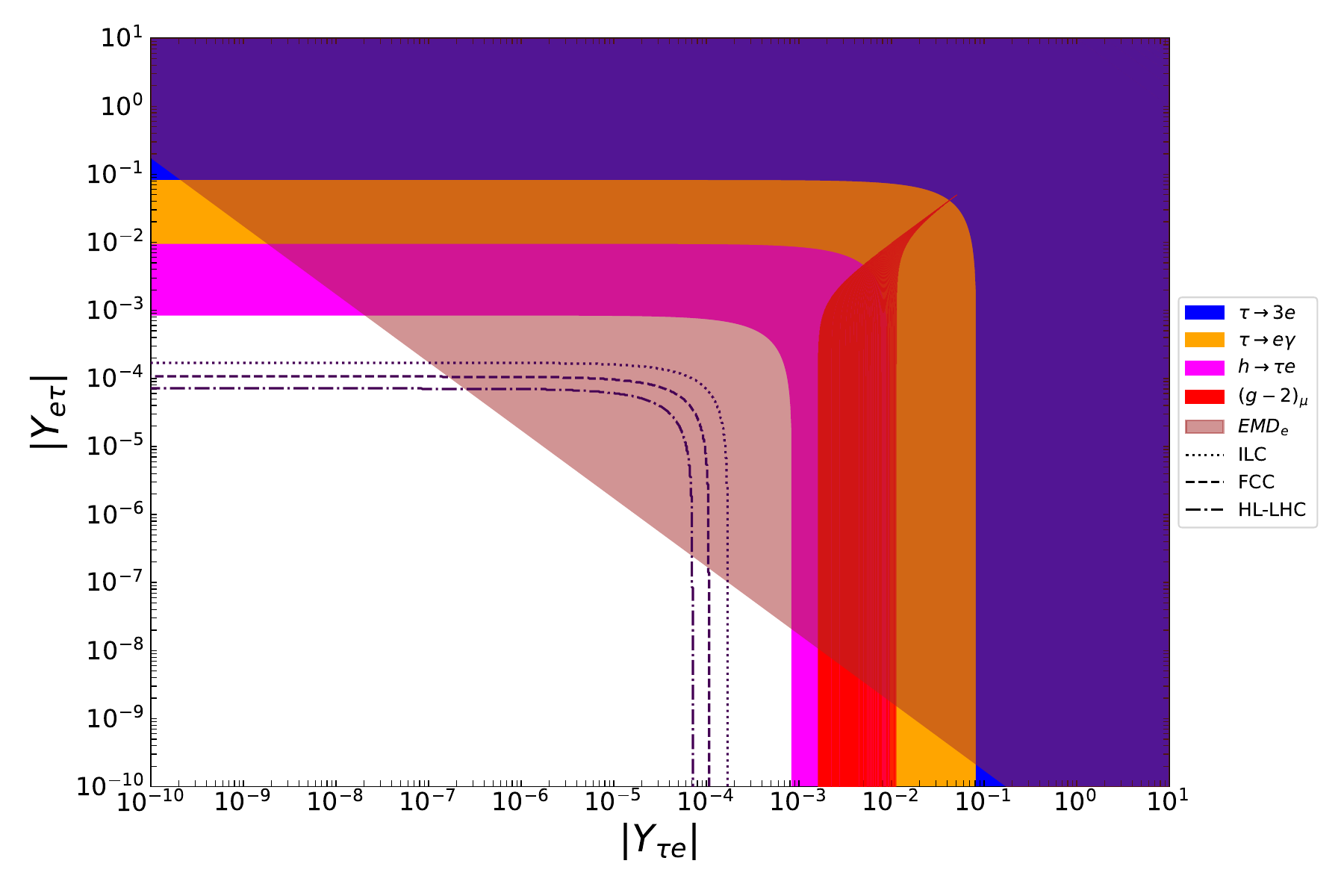}\\
    \end{tabular}
    \includegraphics[width=80mm]{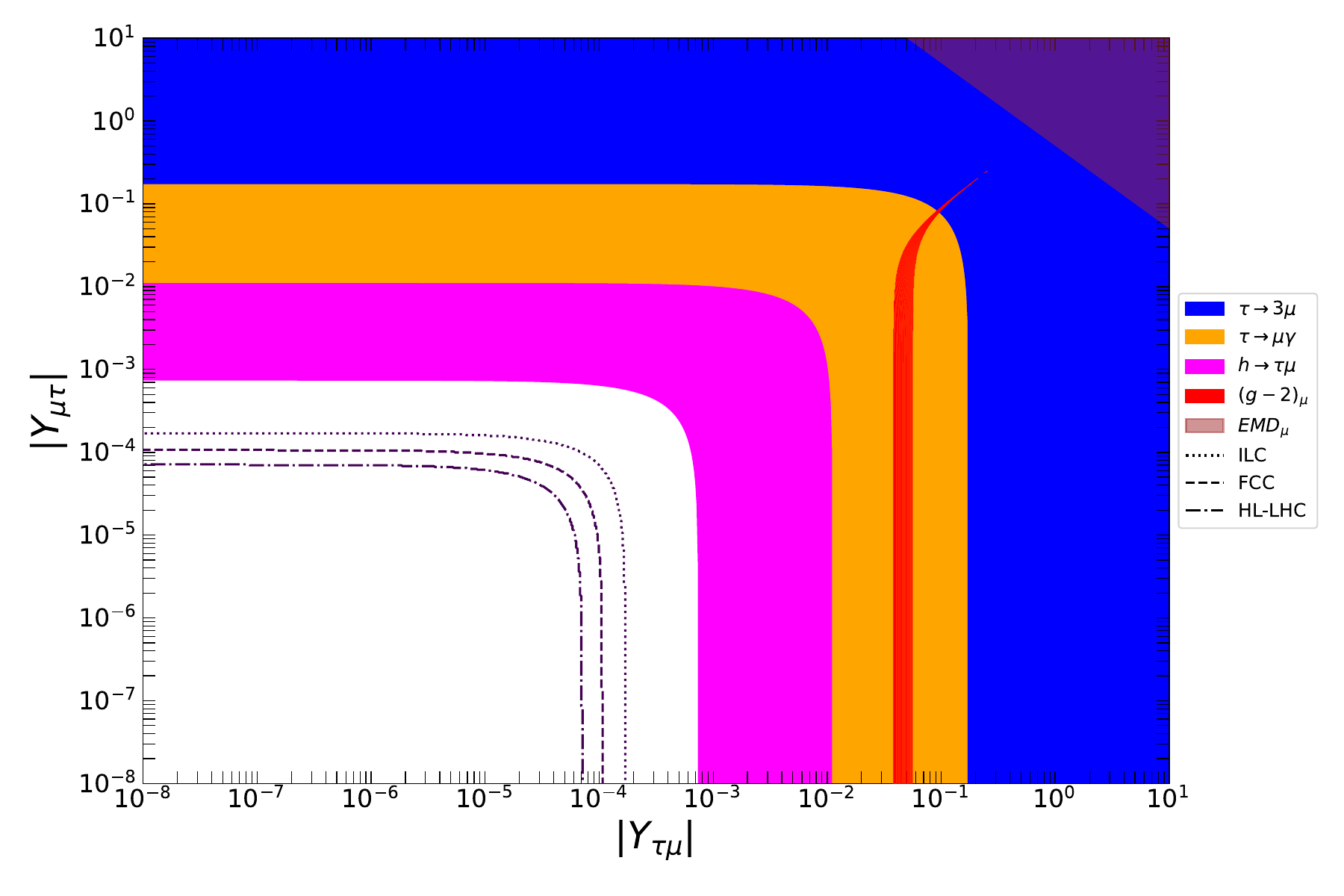}
    \caption{Summary of the experimental limits on the FV couplings of the Higgs to charged leptons. The plots show the excluded region and future projections for the $|Y_{\tau\mu}|-|Y_{\mu\tau}|$, $|Y_{\tau e}|-|Y_{e \tau}|$ and $|Y_{\mu e}|-|Y_{e\mu}|$ parameter space. In addition, we show the region required to explain the $(g-2)_{\mu,e}$ anomalies, where we see that this possibility is already excluded by other bounds. Also notice that the current bounds on $\mu e$ already stronger than all future projections.}
    \label{bounds}
\end{figure}

Bounds are placed on the conversion rate relative to the muon capture rate in the nucleus. According to the SINDRUM II Collaboration, gold yields the strongest bound \cite{SINDRUMII:2006dvw}
\begin{equation}\label{eq:mu_e_cov_rate}
    \frac{\Gamma(\mu \rightarrow e)_{\text{Au}}}{\Gamma_{\text{Capture, Au}}} < 7 \times 10^{-13},
\end{equation}
which given $\Gamma_{\text{Capture, Au}} = 13.07 \times 10^{6}$ $\text{s}^{-1}$ yields $\sqrt{|Y_{\mu e}|^{2} + |Y_{e \mu}|^{2}} < 1.2 \times 10^{-5}$ at $90\%$ CL. On the other hand, the Mu2e experiment is planning on improving this limit by using Aluminum as a target to become $\frac{\Gamma(\mu \rightarrow e)_{\text{Al}}}{\Gamma_{\text{Capture, Al}}} < 1 \times 10^{-16}$ ~\cite{Kargiantoulakis:2019rjm}. Given that $\Gamma_{\text{Capture, Al}} = 0.7054$ $\text{s}^{-1}$, this sets the bound to be $\sqrt{|Y_{\mu e}|^{2} + |Y_{e \mu}|^{2}} < 1.71 \times 10^{-9}$. Finally, we summarize all of these bounds and future projections in Fig.~\ref{bounds}.

\section{$\chi^{2}$ Analysis}
\begin{figure}[t]
    \centering
    \begin{tabular}{cc}
    \includegraphics[width=77mm]{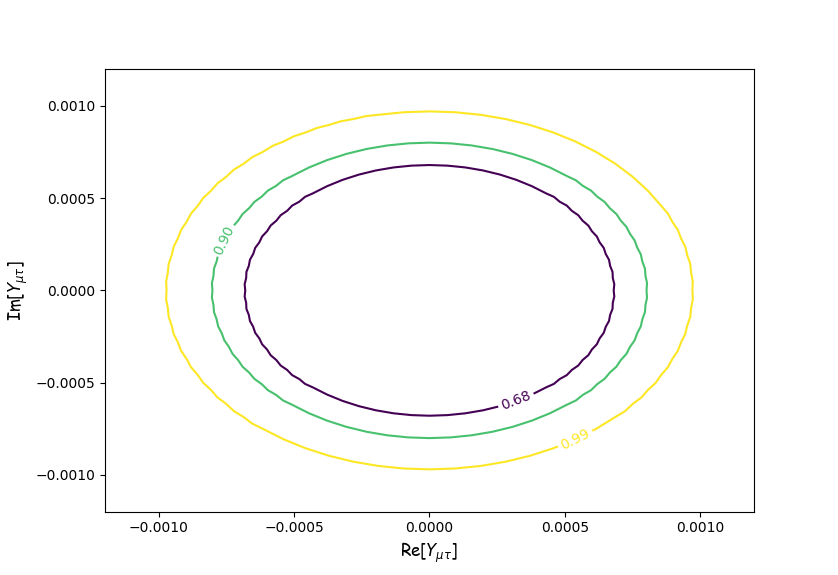} & 
    \includegraphics[width=70mm]{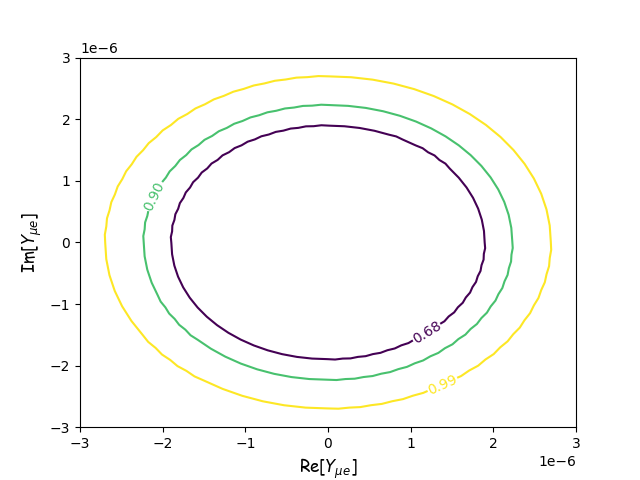}\\
    \end{tabular}
    \caption{The $68\%$, $90\%$ and $99\%$ CL ellipses of $Y_{\mu\tau}$ and $Y_{\mu e}$ are shown in the complex plane for each coupling. }
    \label{region}
\end{figure}

For the sake of this analysis, we assume that the Yukawa matrix is symmetric, i.e., $Y_{\tau \mu}=Y_{\mu \tau}$, $Y_{e \mu}=Y_{\mu e}$ and $Y_{\tau e}=Y_{e \tau}$, and we assume these couplings to be complex $Y_{ij} = |Y_{ij}|e^{\delta_{ij}}$. We perform a $\chi^2$ analysis on the direct and indirect measurements listed in Table~\ref{tab:1} to evaluate the values of FV Yukawa couplings. The $\chi^2$ function is defined as
\begin{equation}
\chi^2 (Y_{ij}) = \sum_{\rm all\, obs.} \frac{\left(O^{\rm th}(Y_{ij}) - O^{\rm exp}(Y_{ij})\right)^2}{\sigma^2_{\rm exp}},
\end{equation}
where $O^{\rm th}$ is theoretical prediction, $O^{\rm exp}$ is the corresponding experimental central value, and $\sigma_{\rm exp}$ is the experimental error for each observable. In order to include an upper limit in the fit, we convert it to a central value plus an error so that we can reproduce the upper limit value at a levels of $1.645\sigma$ (or $90\%$ CL) and $2\sigma$ (or $95\%$ CL). For a given Yukawa coupling, the most likely value is obtained by minimizing $\chi^2$. For this minimization, we use the {\tt MINUIT} library~\cite{James:1975dr,James:1994vla}. The best fit values of the FV Yukawa couplings are listed in table~\ref{bfvalues}. We find that the absolute values of $Y_{\mu\tau}$ and $Y_{e\tau}$ to be of $\sim \mathcal{O}(10^{-3})$ whereas $Y_{e\mu}$ is found to be three orders of magnitude smaller. The phases of these couplings are consistent with zero within the error bars due to the lack of enough constraints on the imaginary part of the couplings. We also find the allowed $68\%$, $90\%$ and $99\%$ CL regions for $Y_{\mu\tau}$ and $Y_{\mu e}${\footnote{For $Y_{\tau e}$, we fail to get the regions beacuse of flat nature of the $\chi^2$ function and a small number of measurements. We hope to get it if we have more constraints on it.}}. These regions are shown in Fig~\ref{region}.

\begin{table}[h!]
\begin{center}
\begin{tabular}{|c|c|c|}
\hline 
$Y_{ij}$ & $|Y_{ij}|$,\,  $\delta_{ij}$ & $\Delta \chi^2$  \\ 
\hline
$Y_{\mu\tau}$ & $(0.40\pm 0.19)\times 10^{-3}$,\, $(0.8\pm 2.0)$ & 4.2 \\
\hline 
$Y_{e\mu}$ & $(1.10\pm 0.50)\times 10^{-6}$,\, $(0.0\pm 2.0)$, & 4.4 \\
\hline
$Y_{e\tau}$ & $(0.46\pm 0.22)\times 10^{-3}$, \, $(0.0\pm 1.7)$& 4.8\\
\hline
\end{tabular}
\end{center}
\caption{Best fit values of the (complex) FV Yukawa couplings $Y_{ij} = |Y_{ij}|e^{\delta_{ij}}$. The phases are in radian. Here $\Delta \chi^2 = \chi^2_{\rm SM} - \chi^2_{\rm bf}$.}
\label{bfvalues}
\end{table}
\section{Matching to the SMEFT}
Having placed upper bounds on the FV Yukawa couplings, it would be interesting to translate these bounds into limits on the scale of NP using the SMEFT. In our calculation, we utilize the so called Warsaw basis \cite{Grzadkowski:2010es} to formulate the higher-dimensional operators. In that basis, there is only one class of operators that contribute to the FV Higgs couplings to leptons, which is the $H^{3}\psi^{2}$, i.e.,
\begin{align}
\mathcal{O}_{\text{dim-6}} & = -\frac{C_{ij}}{\Lambda^{2}}  (H^{\dagger}H)(\overline{\ell}_{i}H e_{j}) + h.c.
\end{align}
This is to be matched to the FV Lagrangian
\begin{equation}\label{eq:FV_Lag}
\mathcal{L} = - Y_{ij}\overline{\ell}_{iL} e_{jR}h + h.c.,
\end{equation}
where $\ell_{i,j} = \{\tau, \mu, e\}$. Plugging the Higgs doublet (in the unitary gauge) leads to the matching condition
\begin{equation}\label{eq:matching}
Y_{ij} = \frac{3v^{2}C_{ij}}{2\sqrt{2}\Lambda^{2}}.
\end{equation}

Now, it is a simple exercise to recast the bounds on the FV Yukawa couplings to bounds on the scale of NP $\Lambda$. This depends on the form of the Yukawa couplings that define the bound. For bounds that have the form $|Y_{ij}|^{2} + |Y_{ji}|^{2} < N^{2}$ and setting $C_{ij} = C_{ji} =1$, one can show that the scale of new physics can be expressed as $\Lambda \geq \frac{\sqrt{3}v}{\sqrt{2N}}$. On the other hand, for bounds of the form $\text{Re}(Y_{ij}Y_{ji})$ or $\text{Im}(Y_{ij}Y_{ji})  < N$, setting $\text{Re(Im)}C_{ij} = \text{Re(Im)}C_{ji} = 1$, one finds that the scale of NP is given by $\Lambda \geq \frac{\sqrt{3}v}{\sqrt{2\sqrt{2N}}}$. Finally, for bounds of the form $|Y_{ij} + Y_{ji}^{*}| < N$ ($M$-$\overline{M}$ oscillation), one can show that the scale of NP is given by $\Lambda \geq \frac{\sqrt{3}v}{(2N)^{1/4}}$. Using the calculated bounds on the Yukawa couplings, the scale of NP corresponding to each is easily calculated. These scales are summarized in Table \ref{tab:1} and we see from the table that the scale of NP ranges from $\mathcal{O}(1) - \mathcal{O}(10^{2})$ TeV depending on the experimental bound. More specifically, and focusing only on the magnitudes of the couplings, the most stringent bounds on the couplings to $\tau \mu$ and $\tau e$ come from direct LHC searches and lead to a scale of NP of $\gtrsim 9$ TeV, whereas the most stringent bound on the coupling to $\mu e$ stems from $\mu \rightarrow e$ conversion and yields to a scale of NP $\gtrsim 87$ TeV. This clearly indicates that any FV in the Higgs sector (assuming it exists) is beyond the reach of any current colliders, although it could be within the reach of some future colliders, such as the FCC and the muon collider. Low energy experiments also constitute suitable venues for searching for FV.

\section{Conclusions}
In this work, we have investigated the present bounds on the FV Higgs couplings to the charged leptons. We have investigated the current bounds from high energy direct searches and from low energy searches, including FV lepton decays, lepton number-violating decays, the magnetic and electric dipole moments of the muon and electron, muonium-antimuonium oscillation, and muon conversion in nuclei. We also utilized the SMEFT to translate these bounds into a lower limit on the scale of NP.

We found that for the Higgs coupling to $\tau \mu$ and $\tau e$, the most stringent bound arises from direct LHC searches at $\mathcal{O}(10^{-3})$, which translates to a scale of NP $\sim 9$ TeV, whereas for the coupling to $\mu e$ we found that the most stringent bounds arise from muon conversion at $\mathcal{O}(10^{-5})$, which leads to a scale of NP of $\sim 87$ TeV. A key difference from the earlier study~\cite{Harnik:2012pb} is that the direct LHC searches for $h \to e\tau$ and $h \to \mu\tau$ now place more stringent bounds on the corresponding Higgs couplings ($\tau e$ and $\tau \mu$) than those obtained from indirect constraints. In contrast, the Higgs coupling to $\mu e$ continues to be most strongly constrained by indirect searches, with the direct limits from $h \to \mu e$ remaining comparatively weaker, consistent with earlier findings.

We also extracted projections on the FV couplings from various future experiments. We performed a $\chi^{2}$ analysis to set predictions of the (complex) FV couplings. We found that the $\chi^{2}$ analysis suggests sizes of the FV couplings of $\sim \mathcal{O}(10^{-3})$ for $Y_{\mu\tau}$ or $Y_{e\tau}$ and $\mathcal{O}(10^{-6})$ for $Y_{e\mu}$, with phases consistent with 0. A key result of this study is that any FV in the Higgs sector (if it exists at all) must be extremely small.

\section{Acknowledgement}
SK would like to acknowledge for financial support through the ANRF National Postdoctoral Fellowship (NPDF) with project grant no PDF/2023/000410. SKV is supported by SERB, DST, Govt. of India Grants MTR/2022/000255 , “Theoretical aspects of some physics beyond standard models”, CRG/2021/007170 “Tiny Effects from Heavy New Physics “and IoE funds from IISC.

\end{document}